# Optical channel waveguides written by high repetition rate femtosecond laser irradiation in Li-Zn fluoroborate glass

Sunil Thomas[a], T. Toney Fernandez[b], Javier Solis[b],*, P.R. Biju[a], N.V. Unnikrishnan[a],*

[a]School of Pure & Applied Physics, Mahatma Gandhi University, Kottayam - 686 560, India

[b]Grupo de Procesado por Laser, Instituto de Optica, CSIC, Serrano 121, 28006 - Madrid, Spain

*E-mail : nvu100@yahoo.com,   j.solis@io.cfmac.csic.es,    Tel. : +91 9745047850

**Abstract**  Low loss, optical channel waveguides have been successfully produced by high repetition rate, femtosecond laser inscription in a Li-Zn fluoroborate glass ($64.9B_2O_3 + 25Li_2O + 10ZnF_2 + 0.1Er_2O_3$). High quality waveguides were produced at 500 kHz, 1 MHz and 2 MHz laser repetition rates, showing a refractive index contrast in the range of $3\text{-}6\times10^{-3}$ depending on various fluences. Dependence of experimental parameters such as average laser power, pulse repetition rate and writing speed on the properties of fabricated waveguides has been discussed. The comparison of optical and compositional characterization techniques evidences an enrichment of B and Zn in the guiding region, while F migrates to the heat diffused region of the written structure.

## 1. Introduction

Fluoroborate glasses fall within a special classification of glasses which are known to be super-ionic conductors or fast-ion conductors [1, 2]. Fluorine ions in glass basically help to evade $OH^-$ groups and usually they are non-bridging ions in a multicomponent glass. But when they bond to the glass matrix, they perform as charge carriers. Such a situation leads fluoroborate glasses to promising applications such as halide sensors, displays, fast switches, fuel cells, optical storages etc. A borosilicate glass is one of the most common glasses finding applications in wide variety ranging from general purposes house hold glasses to astronomical telescope objectives [3]. But in such a composition borate functions as a secondary glass former/intermediate oxide. Boron as the main glass former has received less attention in ultrafast laser based material processing [4-6]. A glass with $B_2O_3$ backbone reduces the linear refractive index considerably, compared to a borosilicate glass. Nevertheless, with the ease of glass formation with variety of glass modifiers, a borate glass could be tuned for its refractive index value from 1.45 to almost 2.3 [3]. Hence, higher $\chi^{(3)}$ values and higher nonlinear refractive indices, $n_2$, are also possible. Such a wide tunability attributes a range of applications to produce various active and passive optical elements. A fluoroborate glass is therefore a tunable and combined workforce of all the above said properties.

In this work, we use an ultrafast laser delivering very high peak power to produce waveguides inside the glass. Hence a glass with lower index is preferable to avoid nonlinear effects which otherwise requires highly complex protocols to bring it under control. Fluoro-borate glass modified with tungsten was recently used to produce waveguides with low repetition rate (100 kHz) femtosecond (fs) laser [4]. Though the laser is classified as low repetition rate, the authors were able to report positive index change with heat accumulation effect due to the very low thermal diffusivity ($2\times10^{-3}$ cm$^2$/s) of the glass. High repetition rate femtosecond laser writing was also carried out in a commercial borate glass but the glass was vastly modified with zinc and bismuth to raise its nonlinear refractive index aiming to fabricate nonlinear optical devices such as optical switches or supercontinuum generators [5]. Current work focuses on producing low loss optical waveguides in fluoroborate glass using high repetition rate femtosecond laser and a preliminary investigation of ion migration in and around the laser affected zone.

## 2. Experimental

The molar composition of the $Er^{3+}$-doped Li-Zn fluoroborate glass preferred to carry out the waveguide inscription was $64.9B_2O_3 + 25Li_2O + 10ZnF_2 + 0.1Er_2O_3$. The glass was prepared by conventional melt quenching technique as follows. About 20 g of the batch composition of the precursors ($H_3BO_3$, $Li_2CO_3$, $ZnF_2$ and $Er_2O_3$) were thoroughly mixed in an agate mortar and melted this homogenous mixture in a platinum crucible using an electric furnace at 900°C for 30 min. The so obtained melt was air quenched by pouring it on to a 300°C preheated brass plate mould and annealed at 300°C about 10 h to remove the thermal stress and strain and is allowed to cool slowly inside the furnace until the room temperature was reached [7]. The glass was polished to very high optical quality using a lapping machine before waveguide writing. The prepared glass has good optical transparency with a dimension of $26 \times 13 \times 5$ mm$^3$.

The channel waveguides in the glass sample were written with high repetition rate fs-laser pulses using a fiber based fs-laser amplifier (Tangerine, Amplitude Systems) with an output beam diameter of 4.3 mm ($1/e^2$), pulse width of 500 fs and operating at a wavelength of 1030 nm [8]. All the optical waveguides in the present glass were written by focusing the laser beam through a 0.4 NA aspheric lens, the diameter ($1/e^2$) of the circularly polarized laser beam before the aspheric lens was 4.3 mm. The pulse repetition rates used to carry out the laser writing were 500 kHz, 1 MHz and 2 MHz with a range of energies (200 – 1000 nJ) and scan speeds (0.05 – 20 mm/s). After laser inscription, the sample was polished to optical quality on both coupling end facets. The morphology of the waveguide cross-sections were recorded by Nikon Eclipse optical microscope in transmission differential interference contrast (DIC) mode. Mode profiling of the waveguides were carried out by launching laser light of wavelength 980 nm and 1640 nm by means of a standard single mode fiber (SMF-28) and image the output mode on to a CCD camera. Modification of the elemental composition at the waveguide cross-section regions was analyzed using wavelength dispersive X-ray spectroscopy (WDS) on JSM-7600F Field Emission Gun-Scanning Electron Microscope (FEG-SEM).

## 3. Results and discussion

Femtosecond laser inscription was carried out at three different repetition rates, 500 kHz, 1 MHz and 2 MHz. High refractive index changes were observed at all repetition rates. Fig. 1 shows the DIC images of the best waveguides, in terms of coupling loss, propagation loss and 980/1640 nm mode confinement, formed using the different repetition rates.

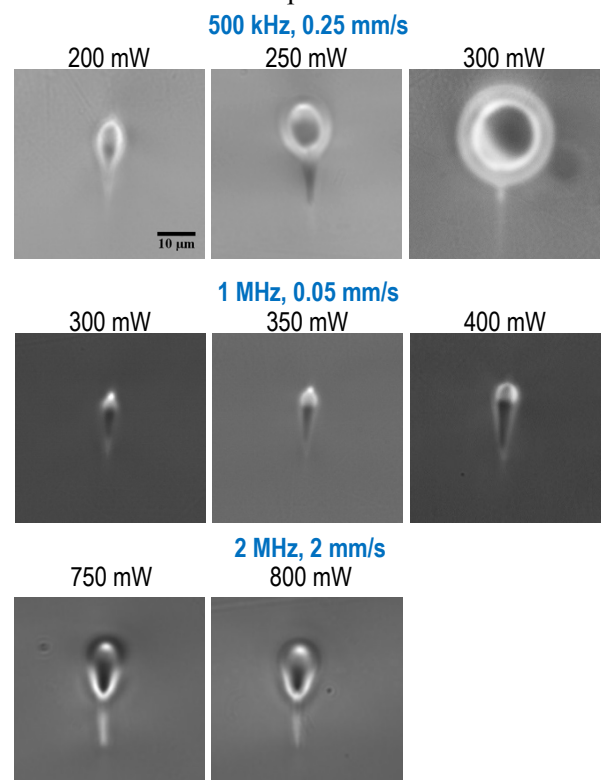

Fig. 1 DIC images of various waveguides written with different fluences 500 kHz, 1 & 2 MHz

All waveguides were profiled for its capability to propagate 980 nm and 1640 nm laser modes (Fig. 2). It was found that waveguides written at 500 kHz and 1 MHz showed single mode propagation at both wavelengths, whereas for 2 MHz, the waveguides showed multimode behavior for 980 nm. For the 1640 nm laser mode propagation, waveguides formed with 2 MHz, 800 mW, 2 mm/s demonstrated a coupling loss of 0.02 dB/facet (when coupled to an SMF-28 fiber) and a propagation loss of 0.4 dB/cm. 500kHz and 1 MHz were found to be more efficient in 980 nm mode propagation demonstrating highly confined guiding, but such waveguides showed slightly higher propagation loss for 1640 nm. Waveguides written at 1 MHz, 300 – 400 mW, 0.05 mm/s demonstrated an average loss of 0.8 dB/cm. The refractive index change ($\Delta n$) of the guiding region has been estimated based on measurements of the NA, for waveguides written with 1MHz to be $\sim 3.5 \times 10^{-3}$. But the maximum refractive index change was observed for 2 MHz written waveguides $\sim 6.6 \times 10^{-3}$. All the waveguides written showed heat accumulation zones [9, 10]. Strong changes in the aspect ratio of the waveguide structure were found when trying to increase the fluence either by increasing the average power or by reducing the speed or repetition rate. Such abrupt changes in the aspect ratio have been shown to be successfully controllable by spatial slit shaping enabling the control of the V-number of the waveguide quite efficiently [5, 11].

Finally ion migration studies by WDS in a SEM were carried out on larger waveguides that were written with 500 kHz at a slightly higher fluence. There are two main reasons behind this approach: first, the glass base composition is formed by very light elements (like boron, lithium and fluorine) and second, the need of having sufficient spatial resolution to track the compositional changes. Conventional energy dispersive X-ray spectroscopy (EDX) has a lower spectral resolution (~150 eV) whereas WDS has a very high spectral resolution (~5 eV). Light elements ($Z < 11$) cannot be analyzed by EDX. Since our glass contains very light elements we used WDS and could analyze compositional changes including boron, whereas lithium is too light even for WDS. The positive index zones formed at 1 MHz (~2.5 µm) and 2 MHz (~1.5 µm) were quite small thus hampering the local composition analysis due to the restriction of measurement time associated with sample drift at high magnification during the WDS spectra acquisition [12]. We decided thus to analyze local compositional changes in structures which are somewhat bigger.

Hence for the first time we report on the ion migration of very light elements like boron and fluorine associated to a waveguide fabricated by an ultrafast laser inscription technique. Fig. 3 shows two waveguides written at 500 kHz, 250 mW (a) 0.25 mm/s and (b) 0.05 mm/s. The single spot spectra acquired from the guiding regions of both waveguides and from an unirradiated region are shown alongside the waveguides.

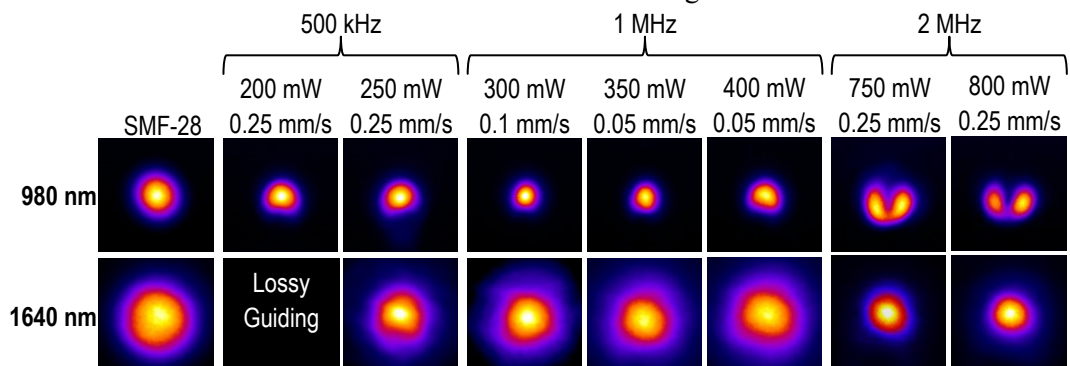

Fig. 2 Propagated modes at 980 and 1640 nm for various waveguides written at

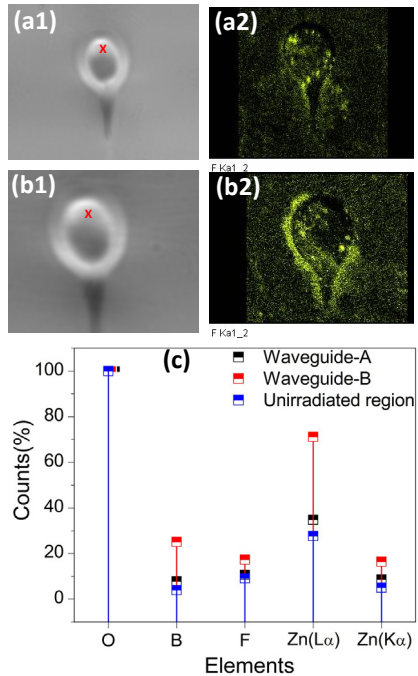

**Fig. 3** Ion migration in two waveguides written at 500 kHz, 250 mW (a) 0.25 mm/s and (b) 0.05 mm/s. a1 & b1 are the DICs, a2 & b2 shows the ion migration of fluorine in respective waveguides and (c) is the comparison of signal counts for both waveguides per element. The spectra were obtained from a spot region as marked in a1 & b1

The results clearly show the enrichment of Zn and B in the guiding region in comparison to the unirradiated sample. This is an expected result as reported before, where multivalent ions move towards the positive index change zone and monovalent ions move towards the negative index change zones [8, 11]. The motion of fluorine, which is monovalent but an anion, is the most interesting result to report. It is evident from the mappings that F enrichment is found in the heat diffusion zones. The motion of monovalent cations towards the defect rich zones reported earlier is well explained via vacancy/interstitial/defect transport in dielectrics. Here $F^-$ anions should in fact be deterred from such defect rich zones [13]. The reason for such a motion could be associated with the heat diffusing out from the heat accumulation region. Fluorine as such is an important element in glass making and for glasses made to utilize especially its active properties. This is the first time the behavior of such a light element is being mapped in waveguide writing.

## 4. Conclusions

Femtosecond laser written optical channel waveguides were fabricated in an $Er^{3+}$-doped Li-Zn fluoroborate glass with different energies and scan conditions. The interplay among the pulse repetition rate, average laser power and the scan speed affects strongly on the waveguide characteristics. The waveguides have revealed good guiding properties like highly confined single mode profiles and good propagation losses which indicate promising potential applications for waveguide amplifiers, lasers and sensors. The guiding of 980 and 1640 nm wavelengths through the waveguides indicates the potential for fabricating waveguide lasers in the near-IR region. The elemental distribution investigation demonstrates migration behavior of the boron, zinc and fluorine ions in the laser irradiated zone. The migration of boron and fluorine within a laser written waveguide is reported for the first time. These results possibly will be supportive for the control of shape and dimension of laser inscribed structures that may show potential applications in integrated photonic devices.

**Acknowledgements** The authors acknowledge SAIF, IIT Bombay for the WDS measurements. One of the authors, Sunil Thomas is also thankful to University Grants Commission (Govt. of India) for the award of RFSMS Fellowship. Two of the authors, P.R. Biju and N.V. Unnikrishnan are also thankful to Department of Science and Technology (Govt. of India) for the financial assistance through DST-PURSE program.